\magnification=\magstep 1
\tolerance 500
\rightline{IASSNS-HEP-96/36}
\vskip 3 true cm
\centerline{\bf Microcanonical Ensemble and Algebra of Conserved Generators}
\centerline{\bf for}
\centerline{\bf Generalized Quantum Dynamics}
\vskip 1 true cm
\centerline{Stephen L.  Adler and L.P. Horwitz\footnote{$^\ast$}{On sabbatical leave from 
School of Physics and Astronomy, Raymond and Beverly Sackler Faculty
of Exact Sciences, Tel Aviv University, Ramat Aviv, Israel, and
Department of Physics, Bar Ilan University, Ramat Gan, Israel.}}
\smallskip
\centerline{School of Natural Sciences, Institute for Advanced Study}
\centerline{Princeton, N.J. 08540}
\vskip 2 true cm
\noindent
{\it Abstract \/}: It has recently been shown, by application of
statistical mechanical methods to determine the canonical ensemble
governing the equilibrium distribution of operator initial values, 
 that complex quantum field theory can emerge as a statistical 
approximation to an underlying generalized quantum dynamics. This
result was obtained by an argument based on a Ward identity analogous to
the equipartition theorem  of classical statistical mechanics. We
construct here a microcanonical ensemble which forms the basis of
this canonical ensemble. This construction enables us to define the
microcanonical entropy and free energy of the field configuration of
the equilibrium distribution and to study the stability of the 
canonical ensemble. We also study the algebraic structure of the
conserved generators from which the microcanonical and canonical
ensembles are constructed, and the flows they induce on the phase space. 

\vfill
\eject
\noindent 
{\bf 1. Introduction}
\par Generalized quantum dynamics$^{1,2}$ is an analytic mechanics on
a symplectic set of operator valued variables, forming an operator
valued phase space $\cal S$.  These variables are defined as the set of
linear transformations\footnote{$^\dagger$}{In general, local (noncommuting) quantum
fields.}
 on an underlying real, complex, or quaternionic
Hilbert space (Hilbert module), for which the postulates of a real,
complex, or quaternionic quantum mechanics are satisfied$^{2-6}$.  The
dynamical (generalized Heisenberg) evolution, or flow, of this phase space is
generated by the total trace Hamiltonian ${\bf H} = {\bf Tr} H$, where 
for any operator ${\rm O}$ we have
$$ \eqalign{
{\bf O}\equiv{\bf Tr} {\rm O} &\equiv {\rm Re Tr} (-1)^F {\rm O} \cr
                              &= {\rm Re} \sum_n \langle n \vert (-1)^F
{\rm O} \vert n \rangle,\cr} \eqno(1.1)$$   
$H$ is a function of the operators $\{ q_r(t) \}, \{p_r(t) \},\ \ {\rm
r}= 1,2, \dots,N$ (realized as a sum of monomials, or a limit of a
sequence of such sums; in the general case of local noncommuting
fields, the index $r$ contains continuous variables), and $(-1)^F$ is a grading operator with
eigenvalue $1(-1)$ for states in the boson (fermion) sector of the
Hilbert space. Operators are called bosonic or fermionic in type if they
commute or anticommute, respectively, with $(-1)^F$; for each $r$, $p_r$ 
and $q_r$ are of the same type.
\par The derivative of a total trace functional with respect to some
operator variation is defined with the help of the cyclic property of
the ${\bf Tr}$ operation.  The variation of any monomial ${\rm O}$
consists of terms of the form ${\rm O}_L \delta x_r {\rm O}_R$, for
$x_r$ one of the $\{q_r \}, \{p_r\}$,
which, under the ${\bf Tr}$ operation, can be brought to the form
$$ \delta {\bf O} = \delta {\bf Tr}{\rm O} = \pm {\bf Tr}
{\rm O}_R {\rm O}_L \delta x_r, $$
so that sums and limits of sums of such monomials permit the
construction of 
$$ \delta {\bf O} = {\bf Tr} \sum_r {\delta {\bf O} \over
\delta x_r } \delta x_r,\eqno(1.2) $$
uniquely defining ${\delta {\bf O}}/{\delta x_r}$.
\par Assuming the existence of a total trace Lagrangian$^{1,2}$ ${\bf
L} = {\bf L} (\{q_r \}, \{ {\dot q}_r \} )$, the variation of the
total trace action
$$ {\bf S} = \int_{-\infty}^\infty  {\bf L} (\{q_r \}, \{ {\dot q}_r
\}) dt   \eqno(1.3) $$
results in the operator Euler-Lagrange equations
$$  {\delta {\bf L} \over {\delta q_r}} - {d \over {dt}} {{\delta {\bf
L}} \over {\delta \dot q_r} } = 0 . \eqno(1.4)$$
As in classical mechanics, the total trace Hamiltonian is defined as a  
Legendre transform, 
$$ {\bf H} = {\bf Tr} \sum_r p_r \dot q_r - {\bf L},\eqno(1.5) $$
where 
$$ p_r = { {\delta {\bf L}} \over {\delta \dot q_r}}.  \eqno(1.6)$$ 
 It then follows from $(1.4)$ that 
$${ {\delta {\bf H}} \over {\delta q_r} } = -\dot p_r  \qquad { {\delta
{\bf H}} \over {\delta p_r}} = \epsilon_r \dot q_r,\eqno(1.7) $$
where $\epsilon_r = 1(-1)$ according to whether $p_r, q_r$ are of 
bosonic (fermionic) type.  
\par Defining the generalized Poisson bracket
$$ \{{\bf A}, {\bf B} \} = {\bf Tr} \sum_r \epsilon_r \left( {\delta
{\bf A} \over \delta q_r} {\delta {\bf B}\over \delta p_r} -
{\delta {\bf B} \over \delta q_r} {\delta {\bf A}\over \delta
p_r}\right), \eqno(1.8a) $$
one sees that
$$ {d{\bf A} \over dt } = {\partial {\bf A} \over \partial t} +
\{{\bf A}, {\bf H} \}. \eqno(1.8b) $$
Conversely, if we define 
$$ {\bf x}_s(\eta) = {\bf Tr} (\eta x_s) , \eqno(1.9a)$$
for $\eta$ an arbitrary, constant operator (of the same type as $x_s$, 
which denotes here $q_s$ or $p_s$), then
 $$  {d{\bf x}_s(\eta)\over dt} = {\bf Tr} \sum_r \epsilon_r \left( {\delta
{\bf x}_s(\eta) \over \delta q_r} {\delta {\bf H}\over \delta p_r} -
{\delta {\bf H} \over \delta q_r} {\delta {\bf x}_s(\eta)\over \delta
p_r}\right), \eqno(1.9b)$$
and comparing the coefficients of $\eta$
on both sides, one obtains the Hamilton equations $(1.7)$ as a
consequence of the Poisson bracket relation $(1.8b)$.
\par The Jacobi identity is satisfied by the Poisson bracket of
$(1.8a)$,$^7$ and hence the total trace functionals have many of the
properties of the corresponding quantities in classical mechanics.$^8$
In particular, canonical transformations take the form 
$$\delta {\bf x}_s(\eta)=\{ {\bf x}_s(\eta), {\bf G} \}, \eqno(1.10a)$$
which implies that 
$$\delta p_r= -{\delta {\bf G} \over \delta q_r}~,~~~
\delta q_r=\epsilon_r  {\delta {\bf G} \over \delta p_r}~,\eqno(1.10b)$$
with the generator ${\bf G}$ any total trace functional constructed from
the operator phase space variables.  Time evolution then corresponds to the 
special case ${\bf G}={\bf H} dt$.
\par It has recently been shown by Adler and Millard$^9$ that a
canonical ensemble can be constructed on the phase space ${\cal S}$, reflecting
the equilibrium properties of a system of many degrees of freedom.
Since the operator
$$ \eqalign{{\tilde C} &= \sum_r (\epsilon_r q_r p_r - p_r q_r) \cr
&=\sum_{r,B} [q_r,p_r] - \sum_{r,F} \{q_r,p_r\},\cr}
\eqno(1.11)$$
where the sums are over bosonic and fermionic pairs, respectively, is
conserved under the evolution $(1.7)$ induced by the total trace
Hamiltonian, the canonical ensemble must be constructed taking this
constraint into account. This is done by constructing the conserved
quantity ${\bf Tr}{\tilde \lambda}{\tilde C}$, for some given constant
anti-hermitian operator ${\tilde \lambda}$.
\par In the general case, in the presence of the fermionic sector, 
the graded trace of the Hamiltonian is not bounded from below, and the
partition function may be divergent. When the equations of motion
induced by the  Lagrangian ${\bf L}$ coincide with those induced by
the ungraded total trace of the same Lagrangian,
$$ {\hat {\bf L}} = {\rm Re Tr} L, $$
without the factor $(-1)^F$, the corresponding ungraded total trace
Hamiltonian ${\hat {\bf H}}$ is conserved; it may therefore be
included as a constraint functional in the canonical ensemble, along
with the new conserved quantity ${\hat{\bf Tr}}
{\hat{\tilde \lambda}}{\hat{\tilde C}}$ (see Appendices 0 and C of ref.
9),
 where
$$ \eqalign{{\hat{\tilde C}} &= \sum_r [q_r, p_r]\cr
 &= \sum_{r,B} [q_r,p_r] + \sum_{r,F} [q_r,p_r] .\cr}
\eqno(1.12) $$
   It was argued that the Ward identities derived from the 
canonical ensemble imply that ${\hat{\tilde \lambda}}$ and
${\tilde \lambda}$ are functionally related, so that they may be 
diagonalized in the same basis (Appendix F of ref. 9).
It was then shown that, since the ensemble averages depend only on
${\tilde \lambda}$ and $(-1)^F$, the ensemble average of any operator
must commute with these operators.  Since the ensemble averaged  operator
 $\langle {\tilde C}\rangle_{AV}$ is
anti-self-adjoint, if one furthermore assumes it is completely
degenerate (with eigenvalue $i_{eff} \hbar$), the ensemble average of
the theory then reduces to the usual complex quantum field theory.
  In this paper, we construct a microcanonical ensemble from which
the canonical ensemble of ref. 9 can be obtained following the usual
methods of statistical mechanics.  This construction gives some
insight into the interpretation of the parameters, relates the
canonical and microcanonical entropies, and identitifies the
generalized free energy. It also permits
estimates of the statistical fluctuations admitted by the canonical
ensemble, and error bounds on the Ward identity which we shall treat   
elsewhere.  We give, in this 
framework, a self-consistency proof of the stability of the 
canonical ensemble.  We then go on to discuss the algebraic structure of 
the canonical generators related to the conserved operators $\tilde C$ and 
$\hat {\tilde C}$, and the flows on phase space induced by these generators.

\bigskip
\noindent
{\bf 2. The microcanonical and canonical ensembles}
\smallskip
\par Introducing a complete set of states $\{ \vert n \rangle \}$ in
the underlying Hilbert space, the phase space operators are completely
characterized by their matrix elements $\langle m \vert x_r \vert n
\rangle \equiv (x_r)_{mn}$, which have the form
$$ (x_r)_{mn} = \sum_A (x_r)^A_{mn}e_A , \eqno(2.1)$$
where $A$ takes the values $0,\,1$ for complex Hilbert space,
$0,\,1,\,2,\,3$ for quaternion Hilbert space (technically, a Hilbert module), 
and just the one value $0$
for real Hilbert space, and the $e_A$ are the associated hypercomplex
units (unity, complex, or quaternionic units$^2$). The mathematical
procedures we establish here are applicable to more general Hilbert
modules; arguments are given in ref. 2, however, for restricting our
attention to these three cases, and we shall therefore concentrate on
the real, complex, and quaternionic structures in the examination of
specific properties. The phase space
measure is then defined
as
$$ \eqalign{ d\mu &= \prod_A d\mu^A, \cr
      d\mu^A &\equiv \prod_{r,m,n} d(x_r)^A_{mn},\cr } \eqno(2.2)$$
where redundant factors are omitted according to adjointness
conditions.  The measure defined in this way is invariant under
canonical transformations induced by the generalized Poisson bracket.$^9$
\par We then define the microcanonical ensemble in terms of the set of
states in the underlying Hilbert space which satisfy $\delta$-function
constraints on the values of the two total trace functionals ${\bf H}~,~~
\hat {\bf H}$ and the 
matrix elements of the two conserved
operator quantities $\tilde C~,~~\hat{\tilde C}$ 
discussed in the previous section.  The volume of the
corresponding submanifold in phase space is given by
$$ \eqalign{\Gamma( E, {\hat E}, \tilde \nu, {\hat {\tilde \nu}}) &= \int d\mu\,\delta( E-{\bf
H})\,\delta({\hat E}- {\hat{\bf H}})\cr
& \prod_{n\leq m,A}\, \delta(\nu_{nm}^A
- \langle n \vert (-1)^F {\tilde C} \vert m \rangle^A)\, \delta({\hat 
\nu}_{nm}^A - \langle n \vert {\hat{\tilde C}} \vert m
\rangle^A ),\cr}  \eqno(2.3)$$
where we have used the abbreviations $\tilde \nu \equiv \{\nu_{nm}^A \}$ and
${\hat {\tilde \nu}} \equiv \{{\hat \nu}_{nm}^A \}$ for the parameters in the
arguments on the left hand side. The factor $(-1)^F$ in the term with
${\tilde C}$ is not essential, but convenient in obtaining the 
precise form given in ref. 9 for the canonical distribution.
The entropy associated with this ensemble is  given by
$$ S_{mic}(E, {\hat E}, \tilde {\nu}, {\hat {\tilde \nu}}) = 
\log \, \Gamma(E, {\hat E},\tilde \nu,{\hat{\tilde \nu}}).\eqno(2.4)$$
As we shall see, it is not possible to associate a temperature to
this structure in the usual simple way.
\par The operators ${\tilde C}$ and ${\hat{\tilde C}}$ are defined in
terms of sums over degrees of freedom.  In the context of the
application to quantum field theory, the enumeration of degrees of
freedom includes continuous parameters, corresponding to the measure
space of the fields. These operators may therefore be decomposed into
parts within a certain (large) region of the measure space, which we
denote as $b$, corresponding to what we shall consider as a {\it
bath}, in the sense of statistical mechanics, and within another (small) 
part of the measure space, which we denote as $s$, corresponding to 
what we shall consider as a {\it subsystem}. We shall assume that the 
functionals ${\bf H}$ and ${\hat{\bf H}}$ may also be decomposed
additively into parts associated with $b$ and $s$; this assumption is
equivalent to the presence of interactions in the Hamiltonian or
Lagrangian operators
which are reasonably localized in the measure space of the fields (the
difference in structure between the Lagrangian and Hamiltonian
consists of operators that are explicitly additive), so that the
errors in assuming additivity are of the nature of ``surface terms''.
The constraint parameters may then be considered to be approximately
additive as well, and we may rewrite the microcanonical ensemble as
$$ \eqalign{ \Gamma ( E, {\hat E}, \tilde \nu, {\hat {\tilde \nu}})&= \int\,d\mu_b
\,d\mu_s\,dE_s\,d{\hat E}_s\, (d\nu^s)\,(d{\hat \nu}^s) \cr
&\times \delta (E-E_s-{\bf H}_b)\,\delta(E_s - {\bf H}_s) \delta({\hat
E} - {\hat E}_s - {\hat {\bf H}}_b) \delta ( {\hat E}_s - {\hat {\bf
H}}_s) \cr
&\times \prod_{n\leq m,A} \delta(\nu_{nm}^A - \nu_{nm}^{A,s}
- \langle n \vert (-1)^F {\tilde C}_b \vert m \rangle^A)\,
 \delta(\nu_{nm}^{A,s}
- \langle n \vert (-1)^F {\tilde C}_s \vert m \rangle^A) \cr
&\times \delta({\hat \nu}_{nm}^A - {\hat \nu}_{nm}^{A,s} -
 \langle n \vert {\hat{\tilde C}}_b \vert m\rangle^A)\, \delta({\hat
 \nu}_{nm}^{A,s} - \langle n \vert {\hat{\tilde C}}_s \vert m
\rangle^A). \cr} \eqno(2.5)$$
We recognize the integrations
over $d\mu_s$ and $d\mu_b$ in $(2.5)$ in terms of the corresponding
microcanonical subensembles, for the bath $b$ and subsystem $s$
respectively, i.e., we may write $(2.5)$ as
$$ \Gamma(E, {\hat E}, \tilde \nu, {\hat {\tilde \nu}}) = \int dE_s d{\hat E}_s 
(d\nu^s)(d{\hat
\nu}^s) \, \Gamma_b (E-E_s, {\hat E}-{\hat E}_s,\tilde \nu - \tilde \nu_s, 
{\hat {\tilde \nu}} - {\hat {\tilde \nu}}_s)\, 
\Gamma_s(E_s, {\hat E}_s,\tilde \nu_s, {\hat {\tilde \nu}}_s). \eqno(2.6)$$ 
 \par We now assume that the integrand in $(2.6)$ has a maximum, for
a large number of degrees of freedom, that dominates the integral.
 In the treatment of the statistical mechanics of classical particles,
the number of degrees of freedom generally vastly exceeds the number
of variables controlling the constraint hypersurfaces in the phase
space; in our case, due to the presence of the constraints imposed by
the operators ${\tilde C}$ and ${\hat{\tilde C}}$, there are an
infinite number of variables, and the question of the development of a
significant maximum may be more delicate.  We will demonstrate,
however, that due to the semidefinite form of the autocorrelation 
matrix of the fluctuations, the canonical distribution that we obtain with this assumption
is at least locally stable.
\par Let us, for brevity, define
$$ \xi = \{ \xi_i \} \equiv \{E, {\hat E}, \tilde \nu, {\hat {\tilde \nu}}\},
 \eqno(2.7)$$ 
where the index $i$ refers to the elements of the set of variables, 
so that $(2.6)$ takes the form
$$\Gamma(\Xi)=\int d\xi_s \Gamma_b(\Xi-\xi_s) \Gamma_s(\xi_s), \eqno(2.8a)$$
where $\Xi$ corresponds to the set of total properties for the whole 
ensemble.  
A necessary condition for an extremum in all of the variables at $\xi_s=\bar{\xi}$ 
is then
$${\partial \over \partial \xi} [\Gamma_b(\Xi-\xi) \Gamma_s(\xi)]|_{\bar \xi}=0, \eqno(2.8b)$$
which implies that 
$$ { 1 \over \Gamma_s (\xi)} {\partial \Gamma_s  \over
\partial \xi_i}  (\xi) |_{\bar \xi} = { 1 \over \Gamma_b(\Xi - \xi)}
{\partial  \Gamma_b\over \partial \Xi_i }(\Xi -
\xi)|_{\bar \xi}.
\eqno(2.8c)$$
The logarithmic
derivatives in $(2.8c)$ define a set of quantities analogous to the
(reciprocal) temperature of the usual statistical mechanics, i.e.,
equilibrium-fixing Lagrange parameters common to the bath and the
subsystem.  We write these separately as
$$\eqalign{ \tau &= {\partial \over \partial E } \log \Gamma_s
(\xi)|_{\bar \xi} \cr
            {\hat \tau} &= {\partial \over \partial {\hat E}} \log
\Gamma_s(\xi)|_{\bar \xi} \cr
 \lambda_{nm}^A &= -{\partial \over \partial\nu_{nm}^A } \log
\Gamma_s(\xi)|_{\bar \xi} \cr
{\hat \lambda}_{nm}^A &= -{\partial \over \partial {\hat \nu}_{nm}^A}\log
\Gamma_s(\xi)|_{\bar \xi}. \cr} \eqno(2.9)$$
According to the definition of entropy $(2.4)$, the bath phase space
volume
is given by
$$\eqalign{ \Gamma_b(\Xi-\xi_s) &= e^{S_b(\Xi -\xi_s)} \cr
                              &\cong e^{S_b(\Xi)} \exp \{-\sum_i \xi_{i,s}
{\partial S_b \over \partial \Xi_i } (\Xi)\},\cr} \eqno(2.10)$$
Neglecting the small shift in argument $\Xi \rightarrow \Xi - \xi_s$, it
follows from  $(2.8a-c)$, $(2.9)$, and $(2.10)$ that
$$  \Gamma_b (\Xi-\xi_s) \cong e^{S_b(\Xi)} \exp\{-\tau E_s - {\hat
\tau}{\hat E}_s + \sum_{n \leq m,A} (\nu_{nm}^{A,s} \lambda_{nm}^A +
{\hat \nu}_{nm}^{A,s} {\hat \lambda}_{nm}^A)\}. \eqno(2.11)$$
\par We now return to $(2.6)$, replacing the phase space volume of the
 bath,  $\Gamma_b$, by the approximate form $(2.11)$, and the
subsystem phase space volume $\Gamma_s$ by the phase space integral
over the constraint $\delta$-functions, i.e. (we use the equality
henceforth, although it should be understood that we have included
just the dominant contribution), 
$$ \eqalign{\Gamma (\Xi) &= \int d\mu_s dE_s d{\hat E}_s (d\nu^s)(d{\hat
\nu}^s) \delta(E_s - {\bf H}_s) \delta ( {\hat E}_s - 
{\hat {\bf H}}_s)  \cr
 &\times \prod_{n\leq m,A} \, \delta(\nu_{nm}^{A,s}
- \langle n \vert (-1)^F {\tilde C}_s \vert m \rangle^A)\,  \delta({\hat
 \nu}_{nm}^{A,s} - \langle n \vert {\hat{\tilde C}}_s \vert m
\rangle^A) \cr
 &\times e^{S_b(\Xi)} \exp\{-\tau E_s - {\hat
\tau}{\hat E}_s + \sum_{n \leq m,A} (\nu_{nm}^{A,s} \lambda_{nm}^A +
{\hat \nu}_{nm}^{A,s} {\hat \lambda}_{nm}^A)\}. \cr}  \eqno(2.12)$$
\par Carrying out the integrals over the parameters, the
$\delta$-functions imply the replacement of the parameters
$E_s$,${\hat E}_s$, $\nu_{nm}^{A,s}$, ${\hat \nu}_{nm}^{A,s}$ in the
exponent by the corresponding phase space quantities.  For the product
$$ \lambda_{nm}^A \langle n \vert (-1)^F {\tilde C}_s \vert m
\rangle^A, \eqno(2.13)$$
we note that the anti-self-adjoint property of ${\tilde C}_s$ implies
that
$$ \langle n \vert (-1)^F{\tilde C}_s \vert m \rangle = - \langle m
\vert (-1)^F
{\tilde C}_s \vert n  \rangle ^* , \eqno(2.14)$$
with $^*$ denoting conjugation of the hypercomplex units, so that
$$ \eqalign{\langle n \vert (-1)^F{\tilde C}_s \vert m \rangle ^0 &= -
\langle m \vert (-1)^F
{\tilde C}_s \vert n \rangle ^0 ,  \cr
\langle n \vert (-1)^F{\tilde C}_s \vert m \rangle^A &= \langle m
\vert (-1)^F
{\tilde C}_s \vert n \rangle^A,~~~ A \neq 0, \cr } \eqno(2.15)$$
for all three cases of real, complex,
 or quaternionic Hilbert spaces.  Thus we have 
$$ {\rm Re} \lambda_{nm} \langle m \vert (-1)^F {\tilde C}_s \vert n
\rangle =  -\sum_A \lambda_{nm}^A \langle n \vert (-1)^F {\tilde C}_s \vert m
\rangle^A. \eqno(2.16a)$$
Defining the operator ${\tilde \lambda}$ for which the matrix elements
are 
$$\eqalign{
\langle n \vert \tilde \lambda \vert n \rangle^A=&\lambda_{nn}^A, \cr
\langle n \vert {\tilde \lambda} \vert m \rangle^A =& {1 \over 2}
\lambda_{nm}^A,~~~n<m,\cr
}\eqno(2.16b)$$ 
we see that the sum over $n \leq m$ of the expression
$(2.16a)$ is ${\bf Tr} {\tilde \lambda}{\tilde C}_s$.  A similar
result holds for the last term of $(2.12)$ (in this case, since we did
not insert the factor $(-1)^F$, we obtain the ${\hat{\bf Tr}}$ functional). 
The volume in phase space is then 
$$ \Gamma (\Xi) = e^{S_b(\Xi)} \int d\mu_s \exp - \{ \tau {\bf H}_s +
{\hat \tau} {\hat {\bf H}}_s + {\bf Tr} {\tilde \lambda} {\tilde
C}_s + {\hat {\bf Tr}} {\hat {\tilde  \lambda}} {\hat {\tilde C}}_s
\}, \eqno(2.17)$$
 so that the normalized canonical distribution function (with the 
 subscripts $s$ removed) is given by 
 $$ \rho = Z^{-1} \exp- \{ \tau {\bf H} +
{\hat \tau} {\hat {\bf H}} + {\bf Tr} {\tilde \lambda} {\tilde
C} + {\hat {\bf Tr}} {\hat {\tilde  \lambda}} {\hat {\tilde C}}
\}, \eqno(2.18)$$
where $$ Z = \int d\mu\, \exp- \{ \tau {\bf H} +
{\hat \tau} {\hat {\bf H}} + {\bf Tr} {\tilde \lambda} {\tilde
C} + {\hat {\bf Tr}} {\hat {\tilde  \lambda}} {\hat {\tilde C}}
\}. \eqno(2.19)$$
\par This formula coincides with that obtained by Adler and
Millard.$^9$ Note that the operators ${\tilde \lambda}$ and
${\hat{\tilde \lambda}}$ appear as an infinite set of inverse
``temperatures'', i.e., equilibrium Lagrange parameters associated both with
the bath and the subsystem, corresponding to the conserved matrix elements of
$(-1)^F {\tilde C}$ and ${\hat{\tilde C}}$. 
\par We finally remark that the microcanonical entropy defined in
$(2.4)$ provides the Jacobian of the transformation from the
integration over the measure of $\cal S$ in $(2.19)$ to an integral
over the parameters defining the microcanonical shells. To see this,
we rewrite $(2.19)$ as 
$$ \eqalign {Z &= \int d\mu dE d{\hat E} (d\nu)(d\hat{\nu})\delta( E-{\bf
H})\,\delta({\hat E}- {\hat{\bf H}})\,\cr
 &\times \prod_{n\leq m,A}\,
 \delta(\nu_{nm}^A
- \langle n \vert (-1)^F {\tilde C} \vert m \rangle^A)\, \delta({\hat 
{\nu}}_{nm}^A - \langle n \vert {\hat{\tilde C}} \vert m 
\rangle^A )\cr
& \times  \exp -\{ \tau E +
{\hat \tau} {\hat E} + {\bf Tr} \tilde \lambda \tilde \nu
+ {\hat {\bf Tr}} \hat {\tilde \lambda} \hat {\tilde \nu}
\}, \cr} \eqno(2.20a)$$
where we have defined the anti-self-adjoint 
parametric operators $\tilde \nu$ and $\hat
{\tilde \nu}$ by  
$$\eqalign{
\nu_{nm}^A=&\langle n \vert (-1)^F \tilde \nu \vert m \rangle^A, \cr
\hat \nu_{nm}^A=&\langle n \vert \hat{\tilde \nu} \vert m \rangle^A. \cr
}\eqno(2.20b)$$
The phase
space integration over the $\delta$-function
factors reproduces the volume of the microcanonical shell associated
with these parameters, i.e, the exponential of the microcanonical
entropy, so that the partition function can be written as
$$ Z = \int dE d{\hat E} (d{\nu})(d{\hat {\nu}}) e^{S_{mic}(E,
{\hat E}, \tilde \nu, \hat{\tilde \nu})} \exp-\{\tau  E +
{\hat \tau} \hat E + {\bf Tr} {\tilde \lambda} \tilde \nu + 
{\hat {\bf Tr}} {\hat {\tilde  \lambda}} \hat {\tilde \nu}
\}. \eqno(2.21)$$

\bigskip
\noindent
{\bf 3. Stability and thermodynamic relations}
\smallskip

\par In this section, we study the stability of the canonical ensemble
as associated with the dominant contribution to the microcanonical
phase space volume. To this end, we formally define the free
energy $A$ as
the negative of the logarithm of the partition function, 
$$ Z \equiv e^{-A(\tau, {\hat \tau},{\tilde \lambda}, {\hat{\tilde
\lambda}})}, \eqno(3.1)$$
so that $(2.19)$ can be written as
$$ 1 = \int d\mu e^{A(\tau, {\hat \tau},{\tilde \lambda}, {\hat{\tilde
\lambda}})}\exp{- \{ \tau {\bf H} +
{\hat \tau} {\hat {\bf H}} + {\bf Tr} {\tilde \lambda} {\tilde
C} + {\hat {\bf Tr}} {\hat {\tilde  \lambda}} {\hat {\tilde C}}
\}}. \eqno(3.2)$$
Differentiating with respect to\footnote{$^\sharp$}{The
hypercomplex index $A$ should not be confused with the conventional
symbol for the free energy.} $\tau,\, {\hat \tau},$ and the matrix
elements $\lambda_{nm}^A, \, {\hat \lambda}_{nm}^A,$
we obtain (as in Eqs. $(49)$ of ref. 9)
$${\partial A \over \partial \tau} = \langle {\bf H}\rangle_{AV}
, \eqno(3.3)$$
$${\partial A \over \partial {\hat \tau}} = \langle  {\hat{\bf H}}
\rangle_{AV}, \eqno(3.4)$$
and using
$$\eqalign{
{\bf Tr} \tilde \lambda \tilde C=&
-\sum_{n \leq m, A} \lambda_{nm}^A \langle n \vert (-1)^F \tilde C \vert
 m \rangle^A, \cr
\hat{\bf Tr} \hat {\tilde \lambda} \hat {\tilde C}=&
-\sum_{n \leq m, A} {\hat \lambda}_{nm}^A \langle n \vert \hat{\tilde C}
\vert m \rangle^A, \cr
}\eqno(3.5)$$
we find
$${\partial A \over \partial \lambda_{nm}^A} = -\langle\langle n
\vert (-1)^F
{\tilde C} \vert m \rangle^A \rangle_{AV} \equiv -\langle 
C_{nm}^A\rangle_{AV}, \eqno(3.6) $$
$${\partial A \over \partial {\hat\lambda}_{nm}^A} = -\langle\langle n
\vert 
{\hat C} \vert m \rangle^A \rangle_{AV} \equiv -\langle
 {\hat C}_{nm}^A\rangle_{AV}.  \eqno(3.7)$$

\par We now consider the identity
$$ 0 = \int d\mu\, ({\bf H} - \langle {\bf H} \rangle_{AV})e^{A(\tau,
 {\hat \tau},{\tilde \lambda}, {\hat{\tilde
\lambda}})}
\exp{- \{\tau {\bf H} +
{\hat \tau} {\hat {\bf H}} + {\bf Tr} {\tilde \lambda} {\tilde
C} + {\hat {\bf Tr}} {\hat {\tilde  \lambda}} {\hat {\tilde C}}
\}}. \eqno(3.8)$$
Differentiating with respect to $\tau$, one finds
$$ \eqalign{0 = \int d\mu\,\bigl( {\partial A \over \partial \tau} -
{\bf H} \bigr)&({\bf H} - \langle {\bf H} \rangle_{AV})e^{A(\tau,
 {\hat \tau},{\tilde \lambda}, {\hat{\tilde
\lambda}})}\cr
&\times \exp{- \{ \tau {\bf H} +
{\hat \tau} {\hat {\bf H}} + {\bf Tr} {\tilde \lambda} {\tilde
C} + {\hat {\bf Tr}} {\hat {\tilde  \lambda}} {\hat {\tilde C}}
\}} - {\partial \langle {\bf H} \rangle_{AV}
 \over \partial \tau} , \cr}                    \eqno(3.9)$$
so that, from $(3.3)$, we find that (as in ref. 9)
$$ \langle ({\bf H} - \langle {\bf H} \rangle_{AV})^2 \rangle_{AV} =
-{\partial \langle {\bf H} \rangle_{AV} \over \partial \tau} = -
{\partial^2 A \over \partial \tau^2} \geq 0. \eqno(3.10)$$
\par In fact, applying this argument to all of the parameters, we
now show that $A$ is a locally convex function.  With this result, we will
prove the stability of the canonical ensemble.
\par The derivative of $(3.8)$ with respect to ${\hat \tau}$ yields,
using the second of $(3.3)$,
$$ \langle ({\bf H} - \langle {\bf H} \rangle_{AV}) ( {\hat {\bf H}} -
\langle {\hat {\bf H}} \rangle_{AV}) = - {\partial \langle {\bf H}
\rangle_{AV} \over \partial {\hat \tau}} = - {\partial^2 A \over
\partial \tau \partial {\hat \tau}}. \eqno(3.11)$$
In the same way that we obtained $(3.10)$, we also find (using
${\hat{\bf H}} - \langle {\hat{\bf H}}\rangle_{AV}$ as a factor in
the integrand),
$$\langle ({\hat{\bf H}} - \langle {\hat{\bf H}} \rangle_{AV})^2 \rangle_{AV} =
-{\partial \langle {\hat{\bf H}} \rangle_{AV} \over \partial {\hat\tau}} = -
{\partial^2 A \over \partial {\hat\tau}^2} \geq 0. \eqno(3.12)$$
We consider next  the identity 
$$ 0 = \int d\mu\, (C_{nm}^A  - \langle C_{nm}^A \rangle_{AV})e^{A(\tau,
 {\hat \tau},{\tilde \lambda}, {\hat{\tilde
\lambda}})}\exp{- \{\tau {\bf H} +
{\hat \tau} {\hat {\bf H}} + {\bf Tr} {\tilde \lambda} {\tilde
C} + {\hat {\bf Tr}} {\hat {\tilde  \lambda}} {\hat {\tilde C}}
\}}.\eqno(3.13)$$
Differentiating with respect to $\lambda_{n'm'}^B$ and
${\hat\lambda}_{n'm'}^B$, we find 
$$ \eqalign{{\partial^2 A \over \partial \lambda_{nm}^A \partial
\lambda_{n'm'}^B} &= - \langle (C_{nm}^A - \langle C_{nm}^A
\rangle_{AV})\,(C_{n'm'}^B - \langle C_{n'm'}^B \rangle_{AV})
\rangle_{AV}, \cr
{\partial^2 A \over \partial {\hat\lambda}_{nm}^A \partial
\lambda_{n'm'}^B} &= - \langle ({\hat C}_{nm}^A - \langle {\hat C}_{nm}^A
\rangle_{AV})\,( C_{n'm'}^B - \langle C_{n'm'}^B) \rangle_{AV})
\rangle_{AV}, \cr 
{\partial^2 A \over \partial {\hat\lambda}_{nm}^A \partial
{\hat\lambda}_{n'm'}^B} &= - \langle ({\hat C}_{nm}^A - \langle {\hat C}_{nm}^A
\rangle_{AV})\,({\hat C}_{n'm'}^B - \langle {\hat C}_{n'm'}^B) \rangle_{AV})
\rangle_{AV}. \cr} \eqno(3.14)$$
Finally, we differentiate $(3.8)$ with respect to $\lambda_{nm}^A$ and
${\hat\lambda}_{nm}^A$ to obtain
$$ {\partial^2 A \over \partial\tau \partial \lambda_{nm}^A } =
\langle ({\bf H} -
\langle {\bf H} \rangle_{AV})( C_{nm}^A - \langle  C_{nm}^A
\rangle_{AV})\rangle_{AV} \eqno(3.15)$$
and
$${\partial^2 A \over \partial\tau \partial {\hat\lambda}_{nm}^A } =
\langle ( {\bf H} -
\langle  {\bf H} \rangle_{AV})({\hat C}_{nm}^A - \langle {\hat C}_{nm}^A
\rangle_{AV})\rangle_{AV},\eqno(3.16)$$
as well as  the corresponding identity with coefficient ${\hat {\bf H}} -
\langle {\hat{\bf H}} \rangle_{AV}$ to obtain
$${\partial^2 A \over \partial{\hat\tau} \partial \lambda_{nm}^A } =
\langle ({\hat {\bf H}} -
\langle {\hat {\bf H}} \rangle_{AV})( C_{nm}^A - \langle  C_{nm}^A
\rangle_{AV})\rangle_{AV} \eqno(3.17)$$
and
$${\partial^2 A \over \partial{\hat\tau} \partial {\hat \lambda}_{nm}^A } =
\langle ({\hat {\bf H}} -
\langle {\hat {\bf H}} \rangle_{AV})( {\hat C}_{nm}^A - \langle 
{\hat C}_{nm}^A)
\rangle_{AV})\rangle_{AV}. \eqno(3.18)$$
Combining $(3.3)$-$(3.18)$, we find that the Taylor expansion of $A$ through 
second derivatives is given by
$$\eqalign{
&A(\tau+\delta \tau,\hat \tau+ \delta \hat \tau, \tilde \lambda + \delta 
\tilde \lambda , \hat {\tilde \lambda} + \delta \hat{\tilde \lambda}\cr
=&A(\tau,\hat \tau,\tilde \lambda, \hat{\tilde \lambda}) +\delta \tau 
\langle {\bf H} \rangle_{AV} +\delta \hat{\tau} \langle \hat{\bf H} 
\rangle_{AV}-\sum_{n\leq m, A}(\delta \lambda_{nm}^A \langle C_{nm}^A 
\rangle_{AV} + \delta \hat{\lambda}_{nm}^A \langle \hat C_{nm}^A
\rangle_{AV}) \cr
-&{1\over2} \langle [ \delta \tau({\bf H} -\langle {\bf H}\rangle_{AV})
+\delta \hat{\tau} (\hat{\bf H}-\langle \hat{\bf H} \rangle_{AV}) \cr
-&\sum_{m\leq n, A} \delta \lambda_{nm}^A (C_{nm}^A-\langle C_{nm}^A 
\rangle_{AV}) + \delta \hat{\lambda}_{nm}^A (\hat C_{nm}^A - \langle 
\hat C_{nm}^A \rangle_{AV} ) ]^2 \rangle_{AV}; \cr
}\eqno(3.19)$$
the uniform negative sign of the quadratic term in the expansion indicates 
that $A$ is a locally convex function, and shows that the matrix of 
second derivatives of $A$ is negative semidefinite.
\par We now turn to the alternative expression of $(2.21)$ for the partition
function, defined in terms of an integral over the parameters of a
sequence of microcanonical ensembles.  The existence of a
maximum in the integrand which dominates the integration assures the
stability of the canonical ensemble; we now show that $(3.19)$ implies the 
self-consistency of our assumption of a maximum.  
\par  Returning to $(2.21)$, we see that the conditions for a maximum
of the integrand at $\xi = \bar \xi$ are that there be a 
stationary point, i.e., 
that
$$\eqalign{ \tau &= {\partial \over \partial E } S_{mic}
(\xi)|_{\bar \xi}, \cr
 {\hat \tau} &= {\partial \over \partial {\hat E}} S_{mic}(\xi)|_{\bar \xi}, \cr
 \lambda_{nm}^A &= -{\partial \over \partial\nu_{nm}^A }S_{mic}(\xi)|_{\bar \xi} ,\cr
{\hat \lambda}_{nm}^A &= -{\partial \over \partial {\hat \nu}_{nm}^A}S_{mic}(\xi)|_{\bar \xi}, 
\cr} \eqno(3.20)$$
together with the requirement that the integrand should decrease 
in all directions, so that this point corresponds to a maximum.  
To make our demonstration of stability more transparent, let us define
$$ \chi = \{\chi_i\} = \{ \tau, {\hat \tau}, -\lambda_{nm}^A, -{\hat
\lambda}_{nm}^A \}, \eqno(3.21)$$
so that $(3.20)$ takes the form 
$$ \chi_i = {\partial S_{mic} \over \partial \xi_i}|_{\bar \xi},
\eqno(3.22)$$
where the indices $i$ are in the correspondence implied by $(3.20)$, 
together with the requirement that the second derivative matrix 
$${\partial^2 S_{mic}\over \partial \xi_i \partial \xi_j} 
={\partial \chi_i \over \partial \xi_j} \eqno(3.23)$$
should be positive definite.  
But the values of $E$, ${\hat E}$, $\nu_{nm}^A$ and
$\hat{\nu}_{nm}^A $ are equal to ${\bf H}$, ${\hat{\bf H}}$,
 $C_{nm}^A$ and ${\hat C}_{nm}^A$ in the microcanonical ensemble, as
seen from $(2.3)$. If the stationary values are those given by $(3.3-4)$ 
and $(3.6-7)$, 
then we must have 
$$\xi_i={\partial A \over \partial \chi_i},\eqno(3.24)$$
which implies that 
the matrix inverse to the right hand side of $(3.23)$ is given by
$$ {\partial \xi_j \over \partial \chi_i } = {\partial^2 A\over
\partial \chi_i \partial \chi_j}, \eqno(3.25)$$
which we have shown to be a negative semidefinite matrix.  This in turn 
implies that the matrix on the right hand side of $(3.23)$ is negative definite, 
giving the condition needed to assure that the stationary point in $(3.20)$ is indeed
a maximum.
\par Assuming this maximum dominates the integration, then the logarithm
of the integral in $(2.21)$ (up to an additive term which is
relatively small for a large number of degrees of freedom) may be
approximated by
$$ A \cong \tau E + {\hat \tau} {\hat E} + {\bf Tr}{\tilde
\lambda}{\tilde C} + {\hat {\bf Tr}} {\hat{\tilde \lambda}}
{\hat{\tilde C}} - S_{mic}( E,\, {\hat E}, {\tilde C}, \,  {\hat{\tilde
C}}), \eqno(3.26)$$
where the arguments are at the extremal values, giving the analog of the   
standard thermodynamical result  $A = E - TS$ for the free energy. 

\bigskip
\noindent
{\bf 4. The operators ${\tilde {C}}, {\hat{\tilde C}}$ as generators}
\smallskip

\par The microcanonical ensemble is constructed as a set of elements
of $\cal S$, which satisfy a constraint described by the value of
${\bf H}$.  This subset of $\cal S$ is invariant to the flow generated
by ${\bf H}$, where we define the flow induced by a functional 
according to the canonical transformation formulas of $(1.10a)$ and $(1.10b)$.
As we have
remarked above, the space is further restricted by values of ${\hat
{\bf H}}$ and, in the canonical ensemble, the values of ${\bf
Tr}{\tilde \lambda}{\tilde C}$ and ${\hat{\bf Tr}}
{\hat{\tilde \lambda}}{\hat{\tilde C}}$. Since these four quantities
have vanishing Poisson brackets with each other under our present
assumptions,  the flow generated by all of these functionals lies in
the constrained subset of $\cal S$. In constructing the microcanonical
ensemble, we constrain the values of the conserved operators
${\tilde C}, {\hat{\tilde C}}$, i.e., 
we constrain the values of all total trace functionals constructed  
by projection from these operators.  
It is therefore instructive to study the action of general total trace 
functionals projected from $\tilde C$ and $\hat{\tilde C}$  
as generators of canonical transformations on the phase space. 

\par We first remark that it was pointed out in ref. 9 that a canonical
generator of unitary transformations on the basis of the underlying
Hilbert space has the form
$$ {\bf G}_{\tilde f} = - {\bf Tr} \sum_r [\tilde f, p_r] q_r, \eqno(4.1) $$
where ${\tilde f}$ is bosonic.
Using (1.11) and the cyclic properties of ${\bf Tr}$, one sees that
$$\eqalign{{\bf G}_{\tilde f} &= - {\bf Tr}{\tilde f} \sum_r  (p_r q_r-
\epsilon_r q_r p_r) \cr
 &= {\bf Tr} {\tilde f}{\tilde C}. \cr} \eqno(4.2)$$
We thus see that the conserved operator $\tilde C$ has the additional 
role of inducing the action of unitary transformations
on the underlying Hilbert space. 

\par That this action preserves the
algebraic properties of functionals of the type ${\bf G}_{\tilde f}$  can be seen
by computing the Poisson bracket, 
$$ \{{\bf G}_{\tilde f}, {\bf G}_{\tilde g} \} = 
{\bf Tr}\sum_r \epsilon_r \bigl(
{\delta {\bf G}_{\tilde f} \over \delta q_r} {\delta {\bf G}_{\tilde g} \over \delta
p_r} - {\delta {\bf G}_{\tilde g} \over \delta q_r} 
{\delta {\bf G}_{\tilde f} \over \delta
p_r} \bigr). \eqno(4.3a)$$
We use the result that
$$ \eqalign{\delta {\bf G}_{\tilde f} &= {\bf Tr} {\tilde f} \delta {\tilde C} \cr
&= {\bf Tr} \sum_r \{\epsilon_r({\tilde f} q_r -q_r {\tilde f}) \delta p_r 
-  ( {\tilde f} p_r - p_r {\tilde f}) \delta q_r\} \cr} \eqno(4.3b)$$
 to obtain
$$ \eqalign{ {\delta {\bf G}_{\tilde f} \over \delta q_r} &= - [{\tilde
f}, p_r] ,
\cr
{\delta {\bf G}_{\tilde f} \over \delta p_r} &= \epsilon_r [{\tilde f}, q_r],\cr}
\eqno(4.4)$$
and hence, expanding out the  commutators,
$$ \eqalign{\{ {\bf G}_{\tilde f}, {\bf G}_{\tilde g} \} &= 
-{\bf Tr} \sum_r \{ p_r {\tilde
f} q_r {\tilde g}  - p_r {\tilde f} {\tilde g} q_r - {\tilde
f} p_r q_r {\tilde g} + {\tilde f} p_r {\tilde g} q_r \cr
&- p_r {\tilde g} q_r {\tilde f} + p_r {\tilde g} {\tilde f}
q_r + {\tilde g} p_r q_r {\tilde f} - {\tilde g} p_r {\tilde
f} q_r\}.   \cr} \eqno(4.5)$$
The first and last terms on the right cancel under the ${\bf Tr}$, as
do the fourth and fifth. These cancellations do not depend on the
grading under the trace, since they involve only cycling of the
bosonic operators $\tilde f,\tilde g$. The remaining terms can be rearranged to the
form
$$ \eqalign{\{ {\bf G}_{\tilde f}, {\bf G}_{\tilde g} \} &= -{\bf Tr} \sum_r
[{\tilde f}, {\tilde g}] (p_r q_r - \epsilon_r q_r p_r) \cr
&= {\bf Tr} [{\tilde f}, {\tilde g}] {\tilde C} \cr
&={\bf G}_{[\tilde f, \tilde g]}  .\cr
} \eqno(4.6)$$
These relations, corresponding to the group properties of integrated
charges in quantum field theory, can be generalized to a ``local''
algebra.  Defining
$$ {\bf G}_{\tilde f r} = {\bf Tr} {\tilde f}{\tilde C}_r, \eqno(4.7a)$$
where
$$ {\tilde C}_r = \epsilon_r q_rp_r -  p_rq_r, \eqno(4.7b)$$
one obtains in the same way that
$$ \{{\bf G}_{\tilde f r}, {\bf G}_{\tilde g s}\} = \delta_{rs} {\bf G}_{
[{\tilde f}, {\tilde g}] r}. \eqno(4.8)$$

\par In studying the flows induced by conserved operators, we shall
also need the properties of generators projected from $\hat {\tilde C}$.
We therefore define\footnote{$^\diamond$}{In terms of this definition, 
$$\hat{\bf Tr} \hat {\tilde \lambda} \hat {\tilde C}=
{\bf Tr} (-1)^F \hat{\tilde \lambda} \hat{\tilde C}={\bf G}_{(-1)^F \hat
{\tilde \lambda}}.$$} 
$$ {\hat{\bf G}_{\tilde f}} = {\bf Tr} {\tilde f}{\hat{\tilde
C}}.\eqno(4.9)$$
Substituting $(1.12)$, we find that the operator derivatives of $\hat{\bf G}_
{\tilde f}$ with respect to the phase space variables are 
$$ \eqalign{ {\delta \over \delta q_r} {\hat{\bf G}_{\tilde f}}
 &=  - (-1)^F[(-1)^F \tilde f,  p_r] 
 =-(\tilde f p_r -\epsilon_r p_r \tilde f),  \cr
{\delta \over \delta p_r}{\hat {\bf G}_{\tilde f}}
&= (-1)^F [(-1)^F \tilde f, q_r]
=\tilde f q_r-\epsilon_r q_r \tilde f. \cr} \eqno(4.10)$$
Computing Poisson brackets in the same way as above, we find that the 
algebra of the generators ${\bf G}_{\tilde f}$ and 
$\hat{\bf G}_{\tilde f}$ closes, 
$$\eqalign{ 
\{ \hat{\bf G}_{\tilde f}, \hat{\bf G}_{\tilde g} \} = 
&{\bf G}_{[\tilde f, \tilde g]},\cr
\{ \hat{\bf G}_{\tilde f}, {\bf G}_{\tilde g} \}
=&\{ {\bf G}_{\tilde f}, \hat{\bf G}_{\tilde g} \}
=\hat{\bf G}_{[\tilde f, \tilde g]}, \cr
}\eqno(4.11)$$
giving a structure reminiscent of the vector and axial-vector charge 
algebra in quantum field theory.  Just as the vector and axial-vector 
charge algebra can be diagonalized into two independent chiral charge 
algebras, so the algebra of (4.6) and (4.11) can be diagonalized 
into two independent algebras 
$${\bf G}_{\pm \tilde f}={1\over 2} ({\bf G}_{\tilde f} \pm 
\hat{\bf G}_{\tilde f}), \eqno(4.12)$$
which obey the algebra
$$ \eqalign{
\{ {\bf G}_{\pm \tilde f} , {\bf G}_{\pm \tilde g} \}=&
{\bf G}_{\pm [\tilde f, \tilde g]}, \cr
\{ {\bf G}_{+ \tilde f}, {\bf G}_{- \tilde g} \}=&0. \cr
}\eqno(4.13)$$
Defining a ``local'' version of $\hat{\bf G}_{\tilde f}$ by 
$$\hat{\bf G}_{\tilde f r} = {\bf Tr} {\tilde f} \hat{\tilde C}_r, 
\eqno(4.14)$$
where
$$ \hat{\tilde C}_r =  q_rp_r -  p_rq_r, \eqno(4.15)$$
the algebras of $(4.11)$ and $(4.13)$ can be converted to local versions 
analogous to $(4.8)$.

\par We now turn to the flows associated with  ${\bf G}_{\tilde f}$ and 
$\hat{\bf G}_{\tilde f}$ when used as canonical generators.
Beginning with ${\bf G}_{\tilde f}$, we consider its action on the 
functional ${\bf x}_s(\eta)$ defined in $(1.9a)$, for which 
$\delta {\bf x}_s(\eta)={\bf Tr} \eta \delta x_s$.  
Defining a parameter $\gamma$ along the motion generated by
${\bf G}_{\tilde f}$,  we choose $\delta x_s$ as $d x_s / {d \gamma}$, 
so that by $(1.10a)$ we have
$$ d {\bf x}_s(\eta) = \{{\bf x}_s(\eta), {\bf G}_{\tilde f} \} d\gamma 
.\eqno(4.16)$$
Comparing $(1.10b)$ with $(4.4)$ and $(4.16)$ gives  
$$ \eqalign{{d q_s \over {d \gamma}} &= [\tilde f, q_s], \cr
{d p_s \over {d \gamma}} &= [\tilde f,p_s]. \cr} \eqno(4.17)$$
In both the boson and fermion sectors we see that, as a solution of
the differential equations $(4.17)$, ${\bf G}_{\tilde f}$
induces the action of a unitary group generated by ${\tilde f}$,
$$ x_s(\gamma)= e^{{\tilde f}\gamma} x_s(0) e^{-{\tilde f}\gamma}.
\eqno(4.18)$$
The unitary
transformation $(4.18)$ preserves the supremum 
operator norm
$$||x_s||={\rm sup}_{\{|n\rangle\}} {|\langle n|x_s|n \rangle| \over
|\langle n|n\rangle| } ,\eqno(4.19)$$
where the supremum is taken over all states $|n\rangle$ in Hilbert space.
\footnote{$^\ast$}{The spectrum of $x_s$ may be unbounded; the
argument we have given above then applies to all bounded functions of
the $x_s$, for which the operator norm exists.  There is, moreover, a
possibility that in the unbounded case, a phase space operator may be
an eigenfunction of ${\tilde f}$, in the sense that
$[{\tilde f}, x_s] = \sigma_s x_s$ for some real $\sigma_s$.
The transformation $(4.17)$ would then correspond to
dilation, therefore admitting conformal transformations on some subset
of the phase space $\cal S$ ( for which preservation of the operator norm does
not form an obstacle).}
\par We next consider the canonical transformation induced on ${\bf x}_s(\eta)$ 
by the functional $\hat{\bf G}_{\tilde f}$ defined in $(4.9)$.
Introducing a parameter $\hat \gamma$ along the motion generated by 
$\hat{\bf G}_{\tilde f}$, we have in this case by $(1.10a)$,
$$d{\bf x}_s(\eta)=\{ {\bf x}_s(\eta) , \hat{\bf G}_{\tilde f} \}
d \hat \gamma .\eqno(4.20)$$
Comparing $(1.10b)$ with $(4.10)$ and $(4.20)$ gives 
$$ \eqalign{{dq_s \over d\hat{\gamma}} &= \epsilon_s (-1)^F 
[(-1)^F \tilde f, q_s]
=\epsilon_s \tilde f q_s -q_s \tilde f,\cr
{dp_s \over d\hat{\gamma}} &= (-1)^F [(-1)^F \tilde f, p_s ]
=\tilde f p_s - \epsilon_s p_s \tilde f.\cr}\eqno(4.21)$$
For the bosonic sector, $(4.21)$ can be rewritten as 
$$\eqalign{
{d q_s \over d \hat{\gamma} }=&[\tilde f, q_s], \cr
{d p_s \over d \hat{\gamma} }=&[\tilde f, p_s],\cr 
}\eqno(4.22)$$
and can be integrated as a unitary
transformation for both $q_s$ and $p_s$, 
 $$ x_s(\hat{\gamma}) = e^{\tilde f \hat{\gamma}} x_s(0)
e^{-\tilde f \hat{\gamma}}. \eqno(4.23)$$

\par For the fermionic sector, however, the grading index $(-1)^F$ anticommutes 
with $q_s$ and $p_s$ and $\epsilon_s=-1$; consequently, the differential 
equations $(4.21)$ in this case take the form
$$ \eqalign{
{dq_s \over d\hat{\gamma}}&= -\{ \tilde f, q_s \} , \cr
{dp_s \over d\hat{\gamma}} &= \{ \tilde f,  p_s  \} ,\cr
} \eqno(4.24)$$
and involve {\it anticommutators} with the operator $\tilde f$, i.e., 
a graded action.
We note, however, that the total trace Lagrangians for which 
$\hat{\tilde C}$ is conserved are ones in which the fermion fields appear as 
bosonic bilinears of the form $p_r q_s$; for these bilinears,  
and for the reverse ordered bosonic bilinears $q_s p_r$, 
we find from $(4.24)$  that 
$$\eqalign{
{d (p_rq_s) \over d \hat{\gamma}}&=[\tilde f,p_r q_s], \cr
{d (q_sp_r)\over d \hat{\gamma}}&=-[\tilde f, q_sp_r]. \cr
}\eqno(4.25)$$
The solution of these differential equations is the unitary group action 
$$\eqalign{
(p_rq_s)(\hat{\gamma})&=e^{\tilde f \hat{\gamma}}(p_rq_s)(0)
e^{-\tilde f \hat{\gamma}}, \cr
(q_sp_r)(\hat{\gamma})&=e^{-\tilde f \hat{\gamma}} (q_sp_r)(0)
e^{\tilde f \hat{\gamma}}, \cr
}\eqno(4.26)$$
which preserves the supremum operator norm of the bilinears $p_rq_s$ and 
$q_sp_r$.  However, it is easy to see that for fermionic operators, 
the supremum 
operator norm of $(4.19)$ is not preserved by the evolution of $(4.24)$
.\footnote{$^\dagger$} 
{For example, to lowest order in $\delta \hat{\gamma}$, 
for the fermionic coordinate variable $q_s$ we have
$$ (n,q_s(\delta \hat{\gamma})n) = (n, q_s n) - [(n, \tilde f
q_s n) + (n, q_s \tilde f n)] \delta \hat{\gamma} 
+O((\delta \hat{\gamma})^2). $$
Now, letting $n= g +h\delta \hat{\gamma}$, we obtain
$$ \eqalign{(n, q_s(\delta \hat{\gamma}) n) &= (g,q_s g) + [(h,q_s g) +
(g,q_s h) ] \delta \hat{\gamma} \cr &-[ (g, \tilde f
q_s g) + (g, q_s \tilde f g) ]\delta \hat{\gamma} +
O((\delta \hat{\gamma})^2). \cr}  $$
Taking $h= \tilde f g$, we see that
$(g, q_s h) = (g,q_s \tilde fg)$
 cancels the fifth term on the right, but 
$ (h, q_s g) = ( \tilde f g, q_s g) 
= -(g, \tilde f q_s g)  $
does not cancel the fourth.  The action of the diagonalized generators   
defined in
$(4.12)$ is therefore norm preserving on bosonic, but not on fermionic 
operators.}

\par  Finally, it is also useful to define parameters $\gamma_{\pm}$ along 
the flows generated by ${\bf G}_{\pm \tilde f}$ according to 
$$d{\bf x}_s(\eta)=\{ {\bf x}_s(\eta), {\bf G}_{\pm \tilde f} \}
d \gamma_{\pm}, \eqno(4.27a)$$
so that 
$${d x_s \over d \gamma_{\pm} }={1 \over 2} \left( {d x_s \over d \gamma }
\pm {d x_s \over d \hat{\gamma} } \right) .  \eqno(4.27b)$$
Then taking sums and differences of $(4.17)$ and $(4.22),~(4.24)$ we 
find that for bosons (with $x_s$ either $q_s$ or $p_s$), 
$$\eqalign{
{d x_s \over d \gamma_+ }=&[\tilde f, x_s] , \cr
{d x_s \over d \gamma_- }=&0, \cr
}\eqno(4.28a)$$
which integrate to 
$$\eqalign{
x_s(\gamma_+)=&e^{\tilde f \gamma} x_s(0) e^{-\tilde f \gamma}  , \cr
x_s(\gamma_-)=&x_s(0) .\cr
}\eqno(4.28b)$$
Similarly, for fermions we find that 
$$\eqalign{
{d q_s \over d \gamma_+}=&-q_s\tilde f~,~~{d p_s \over d \gamma_+}=\tilde f~p_s , \cr  
{d q_s \over d \gamma_-}=&\tilde f~q_s~,~~~~~{d p_s \over d \gamma_-}=-p_s\tilde f , \cr  
}\eqno(4.29a)$$
which integrate to
$$\eqalign{
q_s(\gamma_+)=&q_s(0)e^{-\tilde f \gamma_+}~,~~
p_s(\gamma_+)=e^{\tilde f \gamma_+} p_s(0), \cr
q_s(\gamma_-)=&e^{\tilde f \gamma_-}q_s(0)~,~~
~~p_s(\gamma_-)=p_s(0)e^{-\tilde f \gamma_-} . \cr
}\eqno(4.29b)$$
This identifies ${\bf G}_{\pm \tilde f}$ as the generators of the one-sided 
unitary transformations acting on the fermions which are discussed in 
refs. 1 and 2.
\bigskip
\centerline{\bf Acknowledgments}
This work was supported in part by the Department of Energy under 
Grant \#DE-FG02-90ER40542.  One of us (LH) wishes to thank Hoi Fung Chau 
and Hoi-Kwong Lo for a discussion. 

\centerline{\bf References}
\item{1.} S.L. Adler, Nuc. Phys. B {\bf 415} (1994) 195.
\item{2.} S.L. Adler, {\it Quaternionic Quantum Mechanics and 
Quantum Fields},  Oxford University Press, New York and Oxford, 1995.
\item{3.} E.C.G. Stueckelberg, Helv. Phys. Acta. {\bf 33} (1960) 727;   
{\bf 34} (1961)  621, 675; {\bf 35} (1962) 673. 
\item{4.} D. Finkelstein, J.M. Jauch, S. Schiminovich, and D. Speiser,
J. Math. Phys. {\bf 3} (1962) 207; {\bf 4} (1963) 788. 
\item{5.} L.P. Horwitz and L.C. Biedenharn, Ann. Phys. {\bf 157} (1984) 432. 
\item{6.} C. Piron, {\it Foundations of Quantum Physics}, W.A. Benjamin,
Reading, MA, 1976.
\item{7.} S.L. Adler, G.V. Bhanot, and J.D. Weckel, J. Math. Phys.
{\bf 35} (1994) 531.
\item{8.} S.L. Adler and Y.-S. Wu, Phys. Rev. {\bf D49} (1994) 6705.
\item{9.} S.L. Adler and A.C. Millard, ``Generalized Quantum Dynamics 
as Pre-Quantum Mechanics'', Nuc. Phys. {\bf B}, in press.

\end
\bye